\documentclass[12pt,titlepage]{article}
\usepackage{epsfig}
\usepackage{lscape}
\usepackage{amsmath,epsf,color,colordvi,shadow,pifont}

\begin{document}

\begin{titlepage}
\begin{center}
{\LARGE {\bf  The Apparent Universe}}
 \\
\vskip 1cm

{\large P. Bin\'etruy\footnote{also at Paris Centre for Cosmological Physics 
(PCCP),binetruy@apc.univ-paris7.fr.} and A. 
Helou\footnote{alexis.helou@apc.univ-paris7.fr}} \\
{\em AstroParticule et Cosmologie, 
Universit\'e Paris Diderot, CNRS, CEA, Observatoire de Paris, Sorbonne Paris 
Cit\'e} \\
{B\^atiment Condorcet, 10, rue Alice Domon et L\'eonie Duquet,\\
F-75205 Paris Cedex 13, France}
\end{center}

\centerline{ {\bf Abstract}}
We exploit the parallel between dynamical black holes and cosmological 
spacetimes to describe the evolution of 
Friedmann-Lema\^itre-Robertson-Walker universes from the point of view of an 
observer in terms of the dynamics of the apparent horizon.
Using the Hayward-Kodama formalism of 
dynamical black holes, we clarify the role of the Clausius relation to derive 
the Friedmann equations for a  universe, 
in the 
spirit of Jacobson's work on the thermodynamics of spacetime. We also show
how dynamics at the horizon naturally leads to the 
quantum-mechanical process of Hawking radiation. We comment on the 
connection of this work with recent ideas to consider our observable 
Universe as a Bose-Einstein condensate and on the corresponding role of vacuum 
energy.     
\indent 

\vfill  
\end{titlepage}
\def\noi{\noindent}
\def\sq{\hbox {\rlap{$\sqcap$}$\sqcup$}}
\def\1{{\rm 1\mskip-4.5mu l} }

\newpage
\pagenumbering{arabic}
Since the early proposal of  't Hooft \cite{'tHooft:1993gx} and Susskind
\cite{Susskind:1994vu}, the striking parallel between the dynamics of our 
observable Universe and the dynamics of black holes has been stressed. This has 
led to the formulation of the holographic principle (see \cite{Bousso:2002ju} 
for a review). This has also a deep relation with the AdS/CFT correspondence
\cite{Maldacena:1997re}. On the side of the gravity theory, two seminal works 
have provided a complementary perspective on the issue. G. Gibbons and S. 
Hawking \cite{Gibbons:1977mu} have shown that the close connection between 
event horizons and thermodynamics which has been found in the case of black 
holes can be extended to cosmological models. And Jacobson 
\cite{Jacobson:1995ab} proved
that the Clausius relation $\delta Q = T dS$, familiar to black hole dynamics, 
applied to local Rindler horizons allows to recover Einstein's equations.

In this article, we study the connection between these results in the context 
of Friedmann-Lema\^itre Robertson-Walker (FLRW) with spherical symmetry.
Because cosmology is dynamics, we are using the formalism of Hayward and Kodama
\cite{Hayward:1997jp,Kodama:1979vn} which was devised to describe dynamical 
black holes. As we will see, the 
notion of apparent horizon, which separates trapped and untrapped surfaces
is central to the study.

Our motivation is to identify which part of our classical Universe is relevant 
for understanding the evolution of the Universe, from our standpoint of 
observers. This is important if we wish to see this classical Universe as 
resulting from some measurement process in the quantum mechanical sense. Or 
more precisely if we want to identify where relevant information accessible to 
us, as observers, lies in spacetime.

In the next Section, we review the formalism that we will use, borrowed from the
study of dynamical black holes, and we apply it to cosmologies of FLRW type.    
In Section \ref{sect:2}, we show how the Clausius relation of thermodynamics
applied to the apparent horizon leads to the dynamical evolution of the 
universe according to the standard laws of FLRW cosmology. In the next Section,
we move from thermodynamics to quantum mechanics (in the semi-classical limit)
and show how Hawking radiation can be understood in this context. Finally in 
Section \ref{sect:4}, we interpret the results presented. 

\section{Black hole dynamics vs cosmic dynamics} 
\label{sect:1}

We recall in this Section the formulation of Hayward \cite{Hayward:1997jp}
and Kodama \cite{Kodama:1979vn} for black hole dynamics, before applying it to 
cosmology. The major advantage for us is that this makes the apparent horizon 
play a natural role (whereas the standard approach focuses on the event 
horizon), 
which is easily adapted to the cosmological considerations that follow. For 
simplicity, we consider only spherically symmetric spacetimes: this is general 
enough to be  able to apprehend the dynamics at play in the apparent horizon. 
We thus consider the following spherically symmetric 4-dimensional metric
(with a signature (-,+,+,+)):
\begin{equation}
\label{sphsymmetric}
ds^2 = \gamma_{ij} (x) dx^i dx^j + R^2(x) d\Omega^2 \ ,
\end{equation}
where the two coordinates $x^i$ ($x^0 = t$, $x^1 = r$) run over the 
radial-temporal plane, and $R$ is a function of $t$ and $r$. 

In this context, the total gravitational energy enclosed in a sphere of 
physical radius $R$ is believed to be the Misner-Sharp energy 
\cite{Misner:1964je}:
\begin{equation}
\label{MS}
E(R) \equiv {R \over 2 G} \left( 1 - \nabla^a R \nabla_a R \right) \ ,
\end{equation}
where $G$ is Newton's constant.

A sphere is (anti)trapped, marginal or normal depending on whether 
$\nabla^a R$ is timelike, null or
spacelike. Hence, the surface defined by the condition
\begin{equation}
\label{apparent}
\nabla^a R \nabla_a R = 0
\end{equation}  
corresponds to the hypersurface limiting trapped from normal surfaces: it is 
the apparent horizon, corresponding at a given time to 
a sphere of radius $R_A(t)$ (solution of 
(\ref{apparent})). We deduce that the gravitational energy (\ref{MS}) enclosed
in the apparent horizon is simply
\begin{equation}
\label{ERA}
E(R_A) = {R_A \over 2 G} \ ,
\end{equation}  
an expression familiar from the Schwarzschild black hole.

In black hole (spherically symmetric) dynamics, one has to replace the 
classical notion of Killing vector familiar to the stationary black hole
case, by the Kodama vector \cite{Kodama:1979vn}:
\begin{equation}
\label{Kodama}
K^a \equiv \epsilon^{ab}_\perp \nabla_b R
\end{equation}
where $\epsilon^{ab}_\perp$ is the $(1+1)$ antisymmetric tensor field in the 
$(t,r)$ plane. The Kodama vector lies in this plane and is divergence free:
\begin{equation}
\label{divK}
\nabla_a K^a = 0 \ .
\end{equation}
It satisfies
\begin{equation}
\label{propK}
K^a \nabla_a R = 0 \ , \ \ \ \ K^a K_a = -\nabla^a R \nabla_a R \ ,
\end{equation}
which allows to express the Misner-Sharp energy (\ref{MS}) in terms of
$K^a K_a$.

A central quantity is the energy flux vector \cite{Hayward:1997jp} 
\begin{equation}
\label{psi}
\psi_a \equiv T_a{}^b \nabla_b R + \omega \nabla_a R \ ,
\end{equation} 
where $T_{ab}$ is the energy-momentum tensor and
\begin{equation}
\label{omega}
\omega \equiv -{1\over 2} T^{ij} \gamma_{ij} \ .
\end{equation}
It measures in particular the departure of the Kodama vector  from being a 
Killing vector (compare with $\nabla_a \xi_b + \nabla_b \xi_a = 0$ for a 
Killing vector $\xi_a$)
\begin{equation}
\label{xitok}
K^a \left( \nabla_a K_b + \nabla_b K_a \right) = 8 \pi G R \psi_b
\end{equation}

The Unified First Law of black hole dynamics \cite{Hayward:1997jp} then reads
\begin{equation}
\label{1stlaw}
\nabla_a E = A \psi_a + \omega \nabla_a V \ ,
\end{equation}
where $A = 4\pi R^2$ and $V= {4\pi \over 3} R^3$ are the usual area and volume 
of the sphere {\em in flat coordinates}. 

The second term in (\ref{1stlaw}) is interpreted as a work done by the change 
of the apparent horizon. In order to better understand the first term, 
and to see in what sense this is the first law of black hole 
dynamics, we must introduce the surface gravity, which is defined through
\cite{Hayward:1997jp}
\begin{equation}
\label{kappa}
K^a \left( \nabla_a K_b - \nabla_b K_a \right) = 2 \kappa \nabla_b R \ .
\end{equation}
Useful expressions are
\begin{equation}
\label{kappa1}
\kappa = G {E \over R^2} - 4 \pi \omega R G \ ,
\end{equation}
and
\begin{equation}
\label{kappa2}
\kappa = {1 \over 2} \ \hbox{div} \ {\bf grad} \ R = {1 \over 2 \sqrt{-\gamma}} 
\partial_i \left[ \sqrt{-\gamma} \gamma^{ij} \nabla_j R \right] \ .
\end{equation}

One can then write the first term in the Unified Law as
\begin{equation}
\label{Apsi}
A \psi_a = {\kappa \over 8\pi G} \nabla_a A + R \nabla_a \left({E \over R}
\right)
\end{equation}
Let us now introduce a vector $t^a$ tangent to the apparent horizon, which is 
thus defined by
\begin{equation}
\label{tangent}
t^a \nabla_a \left[ \nabla^b R \nabla_b R \right] = 0 \ .
\end{equation}
Using the definition of Misner-Sharp energy (\ref{MS}), we see 
that\footnote{This is often incorrectly stated in the literature by saying that 
the term $\nabla_a (E/R)$ vanishes on the apparent horizon.} $t^a 
\nabla_a [E/R] = 0$. Hence, projecting the Unified First Law tangentially to 
the apparent horizon, we obtain, using (\ref{1stlaw}) and (\ref{Apsi}),
\begin{equation}
\label{t1stlaw}
dE \equiv t^a \nabla_a E = {\kappa \over 8 \pi G} dA + \omega dV \ ,
\end{equation}
where $dA = t^a \nabla_a A$ and $dV = t^a \nabla_a V$. We recognize in the
first term the Clausius relation $\delta Q = T dS$ with the entropy $S$ 
identified with the area $A$ of the apparent horizon divided by $4G$, and
\begin{equation}
\label{TBH}
T = {\kappa \over 2 \pi} \ .
\end{equation} 
\vskip .5cm
The first who extended this formalism to cosmology and who stressed the role of 
the apparent horizon in this context were Bak and Rey in 1999
\cite{Bak:1999hd}. This was followed by a rather extended literature (see for 
example \cite{Cai:2006rs,Abreu:2010ru} and references therein).

From now on, we will focus on Friedmann-Lema\^itre-Robertson-Walker (FLRW) 
geometries:
\begin{equation}
\label{FRW}
ds^2 = - dt^2 + a^2(t) {dr^2 \over 1-kr^2} + R^2 d\Omega^2 \ , \ \ \ R = r a(t)
\ .
\end{equation}
Then $\nabla_t R = RH$ and $\nabla_r R = a$, where $H \equiv \dot a /a$ as 
usual. It follows from (\ref{apparent}) that, on the apparent horizon 
$1- kr^2 = R^2 H^2$ and thus 
\begin{equation}
\label{RAFRW}
R_A^2 = {1 \over H^2 + k/a^2} \ . 
\end{equation}
Also, the Misner-Sharp energy (\ref{MS}) reads
\begin{equation}
\label{MSFRW}
E = {R^3 \over 2 G} \left[ H^2 + {k \over a^2}\right] \ .
\end{equation}
We will suppose in what follows that the Universe is filled with a (possibly
composite) perfect fluid of energy  density $\rho$ and pressure $p$. Then the 
Friedmann equation $H^2 + k/a^2 = 8\pi G\rho/3$ leads to the seemingly obvious 
result 
\begin{equation}
\label{obv}
E = {4 \pi \over 3} R^3 \rho \ .
\end{equation}
It has been stressed by many authors that it is surprising to see here the 
flat space volume (even for $k \not = 0$): gravitational energy and the perfect 
fluid energy conspire (together with spherical symmetry) to give this familiar
looking contribution. 

The Kodama vector has the following components:
\begin{equation}
\label{KFRW}
K^t = \sqrt{1-kr^2} \ , \ \ \ K^r = -Hr \sqrt{1-kr^2} \ .
\end{equation}
One may note that on the apparent horizon $K_a = -\nabla_a R$. Also in the
de Sitter limit ($k=0$ and $H$ constant), we have $K_t = -1$ and $K_r = -aHR$:
in other words, the Kodama vector coincides with the dilatational Killing vector
$\xi^a$.

From (\ref{psi}), one obtains the components of the energy flux vector:
\begin{equation}
\label{psiFRW}
\psi_t = - {1 \over 2} (\rho + p) RH \ , \ \ \ \psi_r = {1 \over 2} (\rho + p) a
\ .
\end{equation}
We note that, in the de Sitter limit ($\rho + p = 0$), $\psi_a = 0$, in 
agreement with the fact 
that, in this limit, one should recover the Killing equation from 
(\ref{xitok}). This seems in contradiction with the role played by $\psi_a$ for 
recovering the Clausius relation
(see (\ref{t1stlaw})). We will explain in the next section this apparent 
contradiction.

Finally, we give the explicit form of the surface gravity (\ref{kappa1}) or
(\ref{kappa2}):
\begin{equation}
\label{kappaFRW}
\kappa = - {R \over 2}  \left[ 2 H^2 + \dot H + {k \over a^2} \right] \ .
\end{equation}

\section{A closer look at the dynamics at the apparent horizon}
\label{sect:2}
   
As recalled in the Introduction, Jacobson \cite{Jacobson:1995ab,Jacobson:2003wv}
has shown that one recovers Einstein equations from the Clausius 
relation applied to a local Rindler horizon. More precisely, the fraction 
rate of increase of the area element of a local Rindler horizon with respect to 
the affine parameter $\lambda$ is the expansion $\theta$ of the null congruence 
of horizon generators:
\begin{equation} 
\label{J1}
\delta A =  \int \theta d\lambda dA  \ .
\end{equation}
where $\lambda$ is the affine parameter. The geodesic deviation equation gives 
the Raychaudhuri equation for the expansion (see for example \cite{Wald:1984rg},
chapter 9):
\begin{equation}
\label{Ray}
{d \theta \over d \lambda} = -{1 \over 2} \theta^2 - \sigma^2
- R_{ab} k^a k^b \ ,
\end{equation}
where $\sigma^2 = \sigma_{ab} \sigma^{ab}$ is the squared shear of the 
congruence, $R_{ab}$ the Ricci tensor
and $k^a$ is the (null) affine tangent vector to the congruence. At the local 
Rindler 
horizon, $\theta^2$ and $\sigma^2$ vanish; thus $\theta = - \lambda R_{ab} k^a 
k^b$ and
\begin{equation} 
\label{J2}
\delta A = \left. - \int \lambda R_{ab} k^a k^b d\lambda dA \right|_{{\cal H}} \ .
\end{equation}
The Clausius relation $\delta Q =  T dS$, with $T = \kappa/(2\pi)$ and
$dS = \eta \delta A$ ($\eta$ constant to be determined) thus reads:
\begin{equation} 
\label{J3}
\delta Q = \left. - {\kappa \over 2 \pi} \int \eta \lambda R_{ab} k^a k^b 
d\lambda dA \right|_{{\cal H}} \ .
\end{equation}
On the other hand, introducing the local boost Killing vector $\xi^a$, one may 
write the energy flux through the element of horizon surface $d\Sigma^a = k^a 
d\lambda dA$ as
\begin{equation} 
\label{J4}
\delta Q = \int T_{ab} \xi^a d\Sigma^b = \left. - \kappa \lambda \int 
T_{ab} k^a k^b d\lambda dA \right|_{{\cal H}} \ ,
\end{equation}
where we used $\xi^a = - \kappa \lambda k^a$. Comparing (\ref{J3}) and
(\ref{J4}) we obtain $T_{ab} k^a k^b = \eta R_{ab} k^a k^b/(2\pi)$for all null
$k^a$, which implies $(2\pi / \eta)T_{ab} = R_{ab} + f g_{ab}$ for some function 
$f$, which is easily determined to be $f=-R^a{}_a + \Lambda$. One thus recovers 
Einstein's equations with a cosmological constant $\Lambda$, for $\eta = 
1/(4G)$.

It is essential in Jacobson's derivation to consider the locus of vanishing 
expansion (in his case the local Rindler horizon, since it is instantaneously 
stationary). At a more global level, this is precisely the definition of the 
apparent horizon. One may thus expect that the Clausius relation applied to the 
apparent horizon allows to recover the dynamics of FLRW. This is what we will 
now prove, using the formalism introduced in the previous section. In 
particular,
as we will see, when going from local Rindler horizon to the apparent horizon, 
the Killing vector is replaced by the Kodama vector.

In order to describe the appropriate null geodesic congruences, we define the 
null vectors 
\begin{eqnarray}
\label{k}
k^a &\equiv& {1 \over 2} \left( 1, -{\sqrt{1-kr^2} \over a}, 0, 0 \right) \ , \\
\label{l}
l^a &\equiv& 2 \left( 1, {\sqrt{1-kr^2} \over a}, 0, 0\right) \ .
\end{eqnarray}
The vector $k^a$, tangent to the light rays, is chosen to describe the 
future-oriented ingoing null congruence. And the vector $l^a$ is introduced 
in such a way that $k^a l_a = -2$. In this way, the metric \cite{Bousso:2002ju}
\begin{equation}
\label{induced}
h_{ab} \equiv g_{ab} + {1 \over 2} \left( k_a l_b + k_b l_a \right)
\end{equation}
is the induced metric on the 2-spheres (at fixed $r$ and $t$). Then, the 
expansion is defined as \cite{Wald:1984rg}
\begin{equation}
\label{exp}
\theta = h^{cd} \nabla_c k_d \ ,
\end{equation}
where $h^{cd}$ is the inverse induced metric. It follows that the expansion 
for the future-directed ingoing null geodesic congruence is simply 
\cite{Bak:1999hd}
\begin{equation}
\label{expFRW}
\theta = H  -{1 \over R} \sqrt{1-kr^2} \ .
\end{equation}
We check that the apparent radius (\ref{RAFRW}) corresponds to vanishing 
expansion. We also note that, {\em on the apparent horizon}, the vector $k^a$ 
just introduced is aligned with the Kodama vector $K^a$ in (\ref{KFRW}) (and 
thus with $\nabla^a R$ as well):
\begin{equation}
\label{kK}
\left. k_a \right|_{{\cal H}} = {1 \over 2} (-1,-{a \over HR},0,0)
=  {1 \over 2 HR_A} \left. K_a \right|_{{\cal H}}
\end{equation}
As emphasized above, the apparent radius is time-dependent: we have from 
(\ref{RAFRW})
\begin{equation}
\label{beta}
{2 \dot R_A \over H R_A} = 2 \left( 1 - {H^2 + \dot H \over H^2 + k/a^2} 
\right) \equiv \beta^2 \ .
\end{equation}
The notation $\beta$ is motivated by the fact that, in the (A)dS/CFT 
correspondence, this is nothing but the beta function of the associated 
renormalisation group flow, as we will discuss 
below\footnote{Note that de Sitter i.e. $k=0$ and
$H$ constant, corresponds to a zero of the beta function. 
}. We note that the 
surface gravity (\ref{kappaFRW}) simply reads
\begin{equation}
\label{kappaFRWbeta}
\kappa = - {R \over R_A^2} \left( 1 - {\beta^2 \over 4} \right)
\end{equation}

Finally, we have introduced in (\ref{tangent}) the vector $t^a$ tangent to the 
apparent horizon, which satisfies $t^a \nabla_a\left[E/R\right] = 0$. One 
easily checks that, up to a normalization factor,
\begin{eqnarray}
\nonumber
t^a &=& {1 \over 2} \left(1 ,-{RH \over a}(1- \beta^2/2), 0,0 \right) \\
&=& \left( 1 - {\beta^2 \over 4} \right) k^a + {\beta^2 \over 16} l^a \ .
\label{tkl} 
\end{eqnarray}

We have seen above that the Clausius relation is obtained directly from 
projecting the Unified First Law tangentially to the apparent horizon.
Since we have, {\em on the apparent horizon}, $l^a \psi_a = 0$ as well as the 
identification (\ref{kK}), we obtain from (\ref{psi})
\begin{eqnarray}
\nonumber
\left. t^a \psi_a \right|_{\cal H} &=& \left( 1 - \beta^2/4\right) \left. 
k^a \psi_a \right|_{\cal H} = - \left. \kappa R_A k^a \psi_a \right|_{\cal H}
= - {\kappa \over 2 H} \left. K^a \psi_a \right|_{\cal H} \\
&=&  {\kappa \over 2H} \left.  T_{ab} K^a K^b \right|_{\cal H} \ ,
\label{TKK}
\end{eqnarray}
where we have used (\ref{kappaFRWbeta}) as well as the fact that, {\em on the
apparent horizon}, $K_a = -\nabla_a R$ (both of them being null). We note the 
similarity with (\ref{J4}) where the role of the Killing vector is now played 
by the Kodama vector.

We thus obtain ($A = 4\pi R^2$) 
\begin{equation}
\label{Atpsi}
\left. A t^a \psi_a \right|_{\cal H} = 2 \pi \kappa H R_A^4 (p+ \rho) \ .
\end{equation}
whereas one gets from (\ref{Apsi}):
\begin{equation}
\left. A t^a \psi_a \right|_{\cal H} = {\kappa \over 8 \pi G} \left. t^a 
\nabla_a A \right|_{\cal H} = {\kappa R_A \over G} \left. t^a \nabla_a R 
\right|_{\cal H} = {1 \over 4 G} \beta^2 \kappa H R_A^2 \ .
\end{equation}
Hence
\begin{equation}
\label{dynamic1}
\beta^2 = 8\pi G R_A^2 (p+\rho) \ ,
\end{equation}
or equivalently, using (\ref{beta}),
\begin{equation}
\label{dynamic2}
\dot H - {k \over a^2} = - 4\pi G  (p+\rho) \ .
\end{equation}
We have thus recovered the dynamics of FLRW universes from the Clausius 
relation.

We understand why the flux vector $\psi_a$ vanishes for a de Sitter spacetime
($p +\rho =0$): in that case there is no dynamics at the apparent horizon.
We also see that the dynamics at the apparent horizon is fully described by 
the holographic beta function.  

\section{Hawking radiation at the apparent horizon}
\label{sect:3}

The preceding considerations have been purely thermodynamical. We now proceed 
to consider the quantum mechanical process of Hawking radiation 
\cite{Hayward:2008jq}. Hawking radiation at the apparent horizon has been 
considered by many authors, both for dynamical black holes or cosmological  
universes \cite{Cai:2006rs,Cai:2008gw,Zhu:2008hn,DiCriscienzo:2010zza}. 
We will 
follow here the Hamilton-Jacobi formulation of the Pauli-Wilczek tunneling 
method \cite{Parikh:1999mf}, more precisely the work of 
Ref.\cite{Hayward:2008jq} on black holes.

We start by rewriting the metric (\ref{FRW}) in terms of the retarded $u$ and 
advanced $v$ conformal time coordinates:
\begin{eqnarray}
ds^2 = - a^2\left({u+v\over 2} \right) du dv + R^2 d\Omega^2 &,&
\label{double-null} \\
u \equiv \eta - \chi\ , \ \ \ v \equiv \eta + \chi \ \ &,& d\eta \equiv 
{dt \over a(t)} \ , \   d\chi \equiv {dr \over \sqrt{1-kr^2}} \ .
\label{nullcoord}
\end{eqnarray}
We then move to the outgoing Eddington-Finkelstein coordinates where the metric
takes the following form:
\begin{equation}
\label{outEF}
ds^2 = - e^{2\Psi}C dU^2 - 2 e^\Psi dUdR + R^2 d\Omega^2 \ ,
\end{equation}
with ($du = dU$)
\begin{equation}
\label{null2EF}
e^{2 \Psi} C = a^2 \partial_U v \ , \ \ \ 2 e^{\Psi} = a^2 \partial_R v \ ,
\end{equation}
which requires $ \partial_R v > 0$. 
We have more explicitly
\begin{eqnarray}
\label{C}
C &=& 1 - kr^2 -R^2H^2 = 1 - {2GE\over R} \ ,\\
\label{Psi}
e^\Psi &=& {a \over \sqrt{1-kr^2} + RH} \ .
\end{eqnarray}
Since $U$ is a retarded time, any trapped ($C<0$) or marginal ($C=0$) surface 
is past trapped or marginal, as expected in our cosmological set up.

The Kodama vector is simply $K = e^{-\Psi}\partial_U$ and the surface gravity
(\ref{kappa2}) simply reads
\begin{equation}
\label{kappaEF}
\kappa  = {1 \over 2}\left( \partial_R C + C \partial_R \Psi \right) \ .
\end{equation}
 We see that, close to the apparent horizon, where $C \sim 0$, we have
$\kappa \sim \partial_R C/2$.

The BKW approximation of the tunneling probability along the classically
forbidden trajectory from outside the horizon to inside takes the 
form\footnote{We restore $\hbar$ solely for this equation but set it to $1$ 
everywhere else.}:
\begin{equation}
\label{BKW}
\Gamma \propto \exp \left( - 2 {\hbox{Im} I \over \hbar}\right) \ ,
\end{equation}
where $\hbox{Im} I$ is the imaginary part of the action $I$ on the classical 
trajectory, to leading order in $\hbar$.

We consider a massless scalar field $\phi \equiv \phi_0 \exp \left( i I 
\right)$. The equation of motion is the Hamilton-Jacobi equation:
\begin{equation}
\label{HJ}
g^{ab} \nabla_a I \nabla_b I = 0 \ .
\end{equation}
The classical action $I$ simply reads
\begin{equation}
\label{I}
I = - \int \omega e^\Psi dU + \int k dR \ ,
\end{equation}
where we have introduced the Kodama energy $\omega \equiv - K^a \nabla_a I =  
- e^{-\Psi} \partial _U I$ and $k \equiv \partial_R I$. Then, from (\ref{HJ}),
one obtains
\begin{equation}
k (Ck+2\omega) = 0 \ .
\end{equation}
The solution $k=0$ corresponds to outgoing solutions\footnote{One easily shows 
that, on the apparent horizon, $k \propto \partial_\eta I + \partial_\chi I$. 
Hence $k = 0$ corresponds to $I(\eta - \chi)$ i.e. outgoing solutions.}, 
whereas $k = -2\omega/C$
corresponds to ingoing solutions. Near the apparent horizon, $C \sim (R-R_A) 
\partial_R C \sim 2 \kappa (R-R_A)$. Hence the ingoing solution contributes to 
the imaginary part of the action
\begin{equation}
\label{pole}
\hbox{Im} \ I =  \hbox{Im} \int {-\omega \over \kappa (R-R_A - i0)} dR
= - {\pi \omega \over \kappa} 
\end{equation}
The probability (\ref{BKW}) takes the thermal form $\gamma = \exp (-\omega/T)$
with the temperature 
\begin{equation}
\label{Tcosmo}
T = - {\kappa \over 2 \pi} \ .
\end{equation}
Note the difference of sign with respect to (\ref{TBH}), which is consistent 
with the fact that, in the cosmological set up, the surface gravity $\kappa$ 
is negative (for $p <\rho/3$, see (\ref{kappaFRWbeta})) 
\cite{DiCriscienzo:2010zza}.

We see that, contrary to the black hole case, the relevant modes are ingoing, 
which 
is compatible with the fact that the observer is not at infinity, as in the 
black hole, but at the center ($r=0$): Hawking radiation is aimed at this 
central observer. 
  
It is informative also to work with Kodama time \cite{Abreu:2010ru}.
Indeed, introducing the Kodama time $\tau$ by writing
\begin{equation}
\label{Ktime}
d\tau = dU + {e^{-\Psi} \over C} dR \ ,
\end{equation}
the metric (\ref{outEF}) simply reads:
\begin{equation}
\label{Ktimemetric}
 ds^2 = -e^{2\Psi} C d\tau^2 + {1 \over C} dR^2 + R^2 d\Omega^2 \ .
\end{equation}
In this metric, the Kodama vector is simply $K = e^{-\Psi} \partial_\tau$.
A fiducial observer has a 4-velocity parallel to the Kodama vector
$U^a =(e^{-\Psi}/\sqrt{C},0,0,0)$. The only non-vanishing component of 
its acceleration $A^a =U^b \nabla_b U^a$ is $A^1 = C\partial_R \Psi + 
\frac{1}{2} \partial_R C$. Hence defining $A^2 \equiv A^a A_a$, we have
\begin{equation}
\label{acceleration} 
A = {1 \over \sqrt{C}}\left(C\partial_R \Psi + \frac{1}{2} \partial_R C \right)
\ ,
\end{equation}
The Unruh temperature measured at $R$ is then given by the Kodama energy
$e^{-\Psi}A/(2\pi)$:
\begin{equation}
\label{Unruh}
T_U(R)
= e^{-\Psi(R)} {1 \over 2\pi \sqrt{C(R)}}\left|C\partial_R \Psi + \frac{1}{2} 
\partial_R C \right|(R) \ ,
\end{equation}
and the temperature measured by an observer at the origin is obtained by 
multiplying by the redshift factor $\left|g_{\tau\tau}\right|^{1/2}$:
\begin{equation}
\label{Unruhorigin}
T_{U,0}(R) = {\left|g_{\tau\tau}\right|^{1/2}(R) \over \left|g_{\tau\tau}
\right|^{1/2}(0)} T_U(R)
= {1 \over 2\pi}\left|C\partial_R \Psi + \frac{1}{2} 
\partial_R C \right|(R) \ ,
\end{equation}
In particular, one recovers the Hawking temperature (\ref{Tcosmo})
by going to the apparent horizon where $C=0$:
\begin{equation}
\label{UnruhHawking}
T = T_{U,0}(R=R_A) = {1\over 4\pi} \left| \partial_R C \right|_{\cal H}=
{\left| \kappa\right| \over 2\pi} \ .
\end{equation}

\section{The quantum portrait of our visible Universe}
\label{sect:4}

We conclude by proposing an interpretation of the results above. We have seen 
that a genuine dynamics for the evolution of the 
Universe appears only when one departs from the de Sitter regime. Indeed, the 
formalism that we have used has allowed us to make a precise description of 
this departure in terms of horizon dynamics. The beta function $\beta^2 = 
2 \dot R_A/(HR_A)$ (see (\ref{beta})) is non-vanishing, due to the energy 
content of the Universe:
as we have seen in (\ref{dynamic1}) $\beta^2 = 8\pi G R_A^2 (p+\rho)$.
We note that all forms of energy besides vacuum energy generate a nonzero beta 
function. This function is nothing but the holographic beta function introduced 
in the context of AdS/CFT correspondence \cite{Maldacena:1997re}: it quantifies 
the 
renormalisation group flow associated with departures from conformal symmetry.
On the gravitational side, this means that there is no arrow of time in
the context of de Sitter geometry (a fixed point of the beta function). The 
arrow of time is induced by forms of energy other than vacuum energy.

In \cite{Binetruy:2012kx}, we promoted the idea that the classical Universe that
we observe results from a projection of the quantum state of the Universe 
through the measurement process. Since time is intrinsically associated with 
this measurement process, this implies a departure from conformal symmetry, or 
in other words a nonzero beta function. Our results presented in this paper 
indicate that the 
classical Universe that we should consider is the {\em spacetime} region 
bounded by 
the apparent horizon. We showed in \cite{Binetruy:2012kx}, using a model of the 
classical Universe as a Bose-Einstein condensate (in analogy with black holes 
\cite{Dvali:2011aa}), that such a region filled 
with vacuum energy corresponds
to maximal entropy, and thus maximal probability. Again, the  
process of measurement leads to a departure from such a de Sitter regime. The
renormalization group flow will eventually lead to another fixed point and a 
return to the de Sitter regime.

\begin{figure}[h]
\hskip -4cm
\epsfig{file=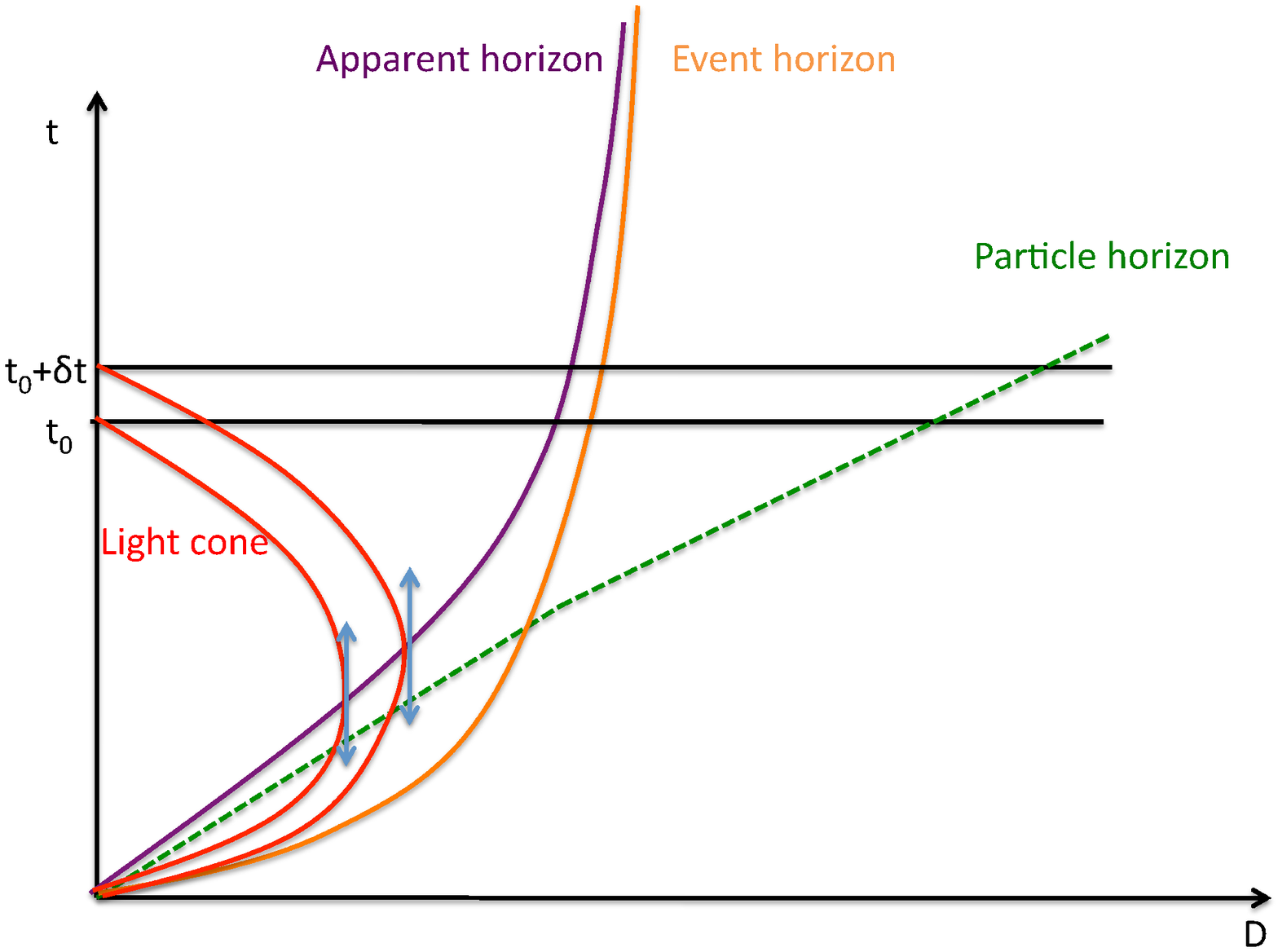,width=18cm,height=15cm}
\caption{The different horizons (apparent in purple, event in orange, particle 
in green) as well as the light cones (in red) for an observer at $t_0$ and 
$t_0 + \delta t$, represented in the plane time versus proper distance, for
a Universe with 70\% vacuum energy and 30\% non-relativistic matter. }
\label{fig1}
\end{figure}

One may be surprised to find that the  apparent horizon plays a central role in 
this picture: we are presently seeing galaxies and stellar objects such as GRBs 
which are beyond the apparent horizon (but within our particle horizon). Let us 
first note that a key assumption in our analysis is spherical symmetry. We must 
thus consider a spherical shell that lies just outside the apparent horizon and
enters it during the time $\delta t$ that we make observations. This is 
represented in Figure \ref{fig1}, in the example of a spatially flat Universe.  
In this case, the apparent horizon is simply the Hubble horizon and includes 
all points that recede from the central observer at a speed smaller than the 
speed of light. Outside the horizon, because of the expansion, photons emitted 
towards the observer 
recede from him (in proper distance) until they are overcome by the 
apparent horizon: from then onwards, they move closer to the observer.
If observation lasts $\delta t$, an infinitesimal  section of spacetime
(limited by the two corresponding light cones, as seen from Figure \ref{fig1})
moves into the apparent horizon. The corresponding information is encoded into 
the Hawking radiation that reaches the observer.

 This work may be pursued along several lines. First, one may relax the 
assumption of spherical symmetry in order to be able to encode 
non-spherical fluctuations. Second, the picture that emerges shows some    
definite differences with the leaking Bose-Einstein model of black hole 
proposed by Dvali and Gomez \cite{Dvali:2011aa,Dvali:2013eja} because Hawking 
radiation leaks into the observable 
patch of the Universe.

Finally, the fertile  parallel between cosmological Universe and black 
hole may in return be an inspiration to understand the dynamics of nonstatic 
black holes, in particular the presence or absence of event horizon
\cite{Hawking:2014tga,Almheiri:2012rt}. 

{\bf Acknowlegments:} We acknowledge the financial support of the UnivEarthS 
Labex program at Sorbonne Paris Cit\'e (ANR-10-LABX-0023 and 
ANR-11-IDEX-0005-02). We wish to thank T. Jacobson, E. Kiritsis, S. Mukhanov 
and G. Veneziano
for useful conversations.

\end{document}